\documentclass{article}
\usepackage{ltwol2ex }

\usepackage{epsfig}

\def\Journal#1#2#3#4{{#1} {\bf #2}, #3 (#4)}

\def\NPB{{\em Nucl. Phys.} B}
\def\NPBs{{\em Nucl. Phys.} B (Proc. Suppl.)}
\def\PLB{{\em Phys. Lett.}  B}
\def\PRL{\em Phys. Rev. Lett.}
\def\PRD{{\em Phys. Rev.} D}
\def\ZPC{{\em Z. Phys.} C}
\def\EPJC{{\em Eur.Phys.J.} C}
\def\IJMP{{\em Int.J.Mod.Phys.}A}

\def\as{\alpha_s}
\def\aspi{\alpha_s/\pi}
\def\qqbar{$\rm q\bar{q}$}
\def\qqbarg{$\rm q\bar{q}g$}
\def\epem{$\rm e^+e^-$}
\def\ppbar{$\rm p\bar{p}$}

\def\be{\begin{equation}}
\def\ee{\end{equation}}
\def\bea{\begin{eqnarray}}
\def\eea{\end{eqnarray}}

\begin{document}

\title{PROPERTIES OF GLUON AND QUARK JETS }

\author{KLAUS HAMACHER}

\address{Fachbereich Physik, Bergische Universit{\"a}t - Gh. Wuppertal,
Gau\ss{}stra\ss{}e 20, 42097 Wuppertal, Germany\\ DELPHI Collaboration\\ E-mail:
Klaus.Hamacher@cern.ch}


\twocolumn[\maketitle\abstracts{ Recent developments and results
on the comparison of gluon to quark jets are discussed. A most
important topic is the introduction of explicit energy scales of
the jets. The scaling violation of the fragmentation function and
the increase of the multiplicity with scale is shown to be
directly proportional to the corresponding gluon or quark colour
factor. The ratio of the hadron multiplicity in gluon to quark jets is
understood to be smaller than the colour factor ratio due to
differences in the fragmentation of the leading quark or gluon.
Novel algorithms to reconstruct the colour portraits or the colour
flow of an event are presented.\\
{\em Talk given at the XXIX ICHEP, Vancouver, July 1998. A
shortened version will appear in the conference proceedings.} }]

\section{Introduction}\label{sec:intro}
QCD contains two types of colour charged fields, quarks and
gluons. The relative coupling strength of the quark gluon vertex
and of the triple gluon vertex is determined by the QCD colour
factors $C_F$ and $C_A$. The values of the colour factors directly
follow from the $SU(3)$ group structure of QCD. Reversely a
precise measurement of the ratio of the couplings uniquely fixes
the gauge group.

A basic prediction thus is that there should be about
$C_A/C_F=2.25$ times more gluon bremsstrahlung in gluon compared
to quark jets \cite{brodsky_gunion}.   Scaling violation of the
fragmentation function \cite{qg_splitt,vanc_scaling} and the
production of soft hadrons \cite{vanc_mult} is
correspondingly stronger in gluon jets.

The comparison of gluon and quark jets thus allows for basic and
intuitive QCD tests. Moreover this comparison offers a direct
handle on the fragmentation, the transition of colour charged
fields into hadrons independent of fragmentation models.

Due to colour confinement there are, however, no free quarks and
gluons. The definition of gluon and quark jets therefore relies on
the analogy to tree level graphs. Corrections of order $\aspi$
have always to be expected. As QCD is a quantum theory also
interference effects are anticipated due to the underlying event
structure and due to the coherence of soft radiation. Both effects
have indeed been observed. Finally effects due to the hadronic
final state should be present. This is, however, part of
fragmentation and what is to be understood. A final
point of complication are the ambiguities introduced by
the algorithms which are used to define the jets.

\section{Experimental Topics and Energy Scales}\label{sec:exp}
The success of gluon and quark jet comparisons at LEP is due to
two topics - the possibility of an unbiased tag of gluon jets and
the better understanding of the (energy) scales underlying jet
evolution.

Gluon jets were originally identified in (2- or 3-fold) symmetric
three-jet ({\em Y} or {\em Mercedes} ) events where two jets could be
identified as (heavy) quark jets using impact parameter tags or
energy ordering (see e.g. \cite{qg1_delphi}).
The remaining  jet is then a gluon jet. A similar
light quark result can be obtained from a mixed
quark/gluon sample from symmetric events by subtracting the gluon.
Recently this technique has also been extended to non-symmetric
topologies \cite{martin_diplom}. In this case one relies on the
quark/gluon composition as predicted by the three-jet matrix
element. The validity of this technique can, however,  be tested
by comparing results obtained with symmetric and non-symmetric
events. This technique strongly improves the available statistic
of gluon jets giving access to a much wider range of energy scales
and allows so to compare the dynamics of gluon and quark jets.

A proper gluon to quark comparison has to be done with jets of a
comparable scale. It turns out that this scale is not just the jet
energy. The phase space for soft radiation is limited by coherence
effects to cones given by the opening angles between the jets
(angular ordering). The relevant scale thus is in general a
product of jet energy $E$ times opening angle $\theta$ i.e. a
transverse momentum. This is what has originally been used in the
so called MLLA (modified leading log approximation) calculations
which were limited to small angles. If also larger opening angles
are included it turns out that the so called hardness
$\kappa=2E\sin\theta/2$ is a better choice \cite{khoze_ochs,durham}.
This definition corresponds to an equivalent CMS energy.
It should however be kept in mind that
in multi-jet events several scales are relevant, in principle.
In so far the usage of $\kappa$ is an approximation which needs to
be verified with data.

\section{Results}\label{sec:results}

\begin{figure*}[t]
\parbox{8.6cm}{ \epsfig{figure=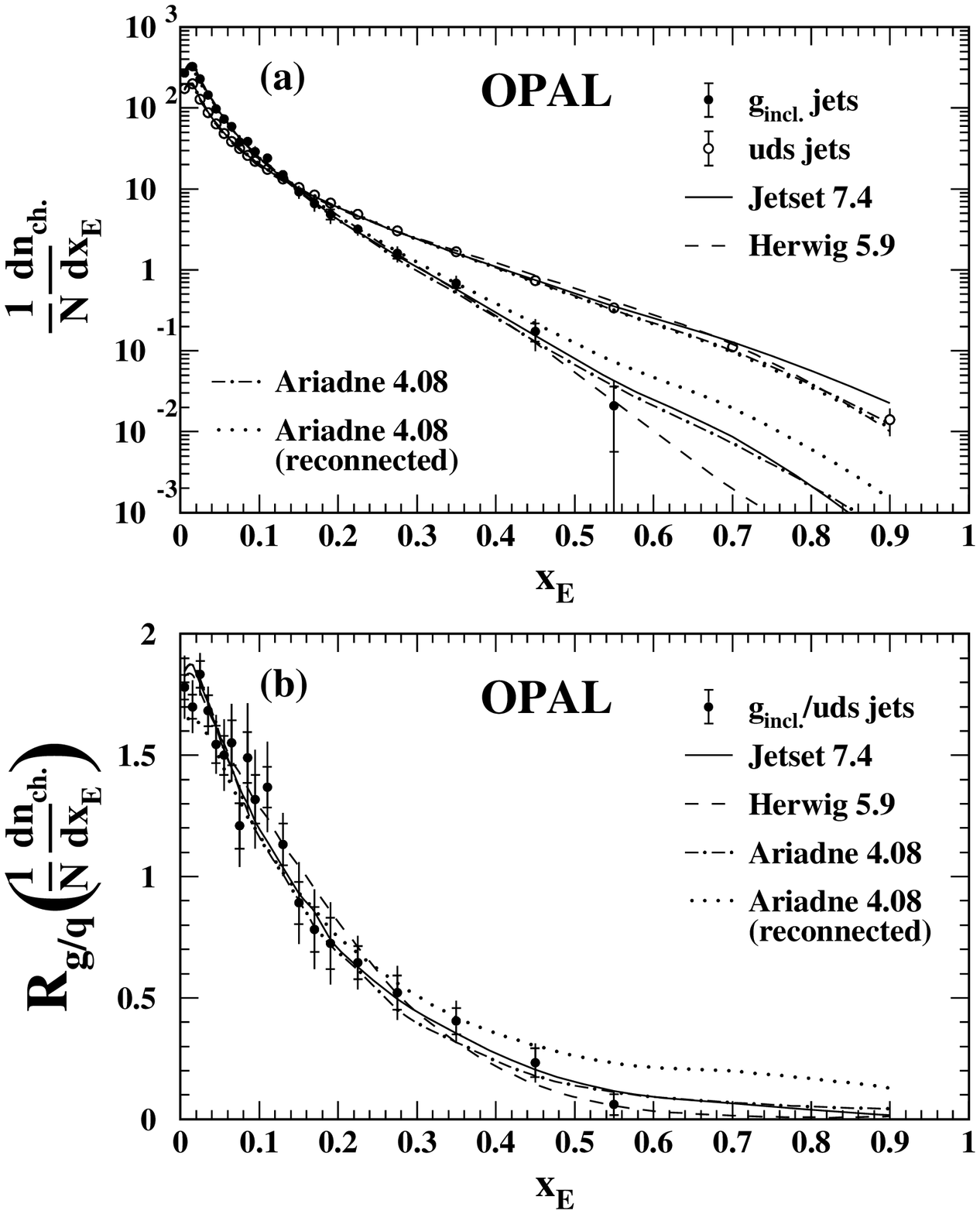,width=8.6cm} }
\hfill
\parbox{8.6cm}{ \epsfig{figure=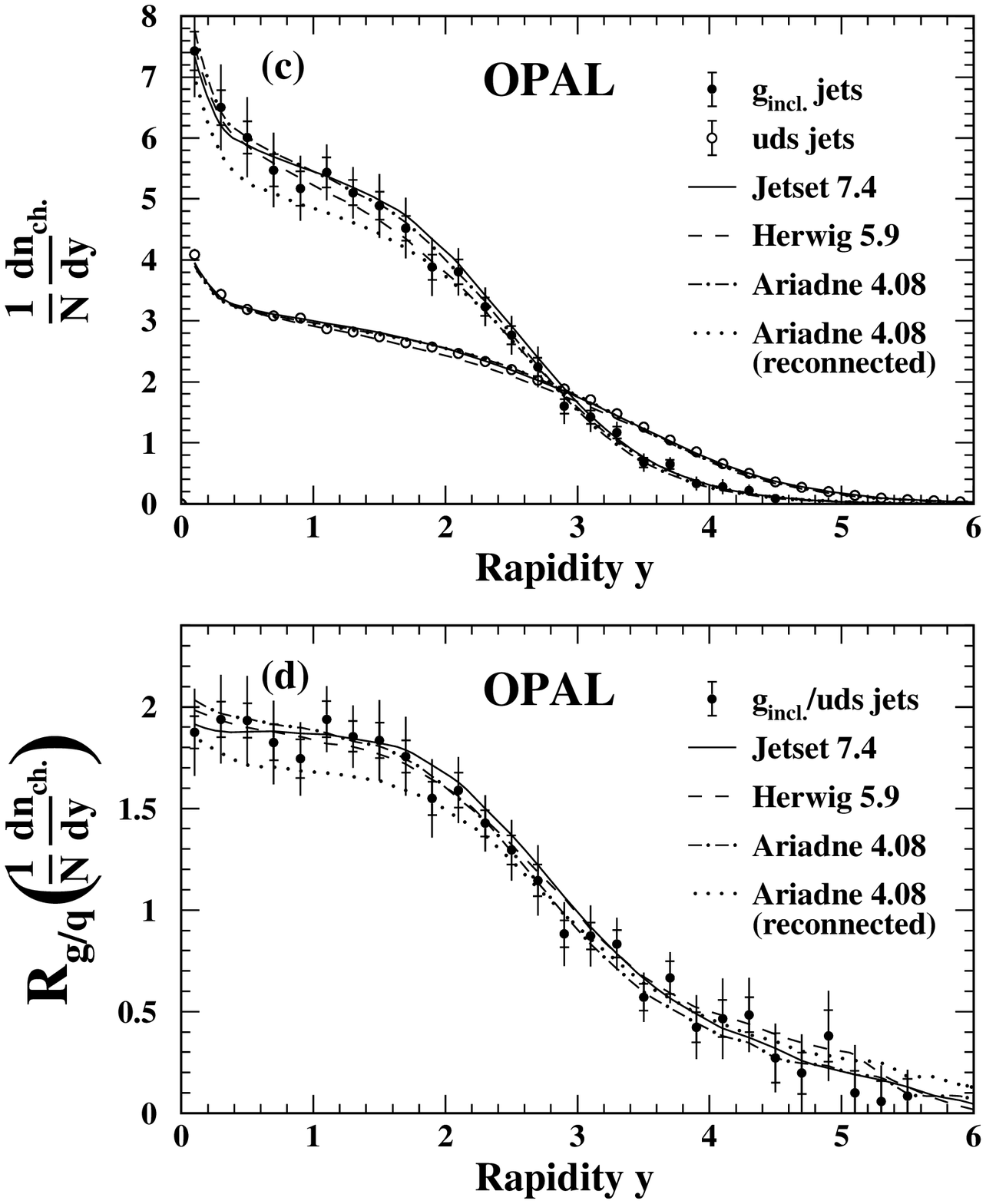,width=8.6cm} }
\center{
\caption{ Gluon and quark fragmentation functions to
charged hadrons as function of the scaled hadron energy $x_E$ (a) and
the rapidity y (c) with respect to the sphericity axis as measured
by OPAL. The lower plots (b,d) show the gluon to quark ratio.}
\label{fig:fragfct}}
\end{figure*}

\subsection{Properties of Gluons at Fixed Scale}\label{subsec:singleg}
So far measurements observed gluon to quark multiplicity ratios
in the range 1. to 1.5 \cite{mult_rat_exp}.
This low ratio has been ascribed to missing energy conservation in the
theoretical calculations \cite{what_opal}.
To minimise this effect as well as
overlap effects between the jets it has been suggested \cite{gary}
to study the most energetic gluons accessible at LEP, in events
where the gluon recoils with respect to the \qqbar\ -pair. The study of these
rare events by OPAL \cite{opal_vanc} yields only 546 gluons from
the total data set taken at the Z resonance. The average gluon
energy is $\approx$42GeV. The assignment of hadrons to the gluon
is made inclusively: all hadrons in the gluon hemisphere of the
event are assigned to the gluon.

This selection
proves to be robust against the usage of different cluster
algorithms \cite{opal_mult}. A comparison of these inclusive
gluon-jets with
identified uds quark jets yields multiplicity distributions with
parameters given in \mbox{Tab. \ref{tab:opal_mult}.} Skewness, kurtosis
and multiplicity differ significantly between the gluon and quark
distributions. If the slightly higher energy of the quark jets is
considered by a Monte Carlo correction, even for these high
energy gluons the multiplicity ratio still attains values
much smaller than the simple QCD expectation
\cite{opal_vanc}:
\begin{eqnarray}
\frac{N_{ch}^{gluon}}{N_{ch}^{quark}} = 1.509 \pm 0.022 \pm 0.046
\label{eqn:opalmultrat}
\end{eqnarray}

\begin{table}
\begin{center}
\caption{Parameters of the gluon and quark multiplicity
 distributions measured by OPAL.
$\left< N_{ch} \right>$ is the mean charged multiplicity,
 $D$ is the dispersion,  $\gamma$ is the skewness and $c$ is the kurtosis.
 }\label{tab:opal_mult}
\vspace{0.2cm}
\begin{tabular}{|c|c|c|}
\hline
                        & Gluon                       & Quark \\
\hline
$\left< N_{ch} \right>$ & 14.32 $\pm$ 0.23 $\pm$ 0.40 & 10.10 $\pm$ 0.01 $\pm$ 0.18 \\
$D$                     &  4.37 $\pm$ 0.19 $\pm$ 0.26 &  4.30 $\pm$ 0.01 $\pm$ 0.10 \\
$\gamma$                &  0.38 $\pm$ 0.13 $\pm$ 0.18 &  0.82 $\pm$ 0.01 $\pm$ 0.04 \\
$c$                     &  0.18 $\pm$ 0.34 $\pm$ 0.30 &  0.98 $\pm$ 0.03 $\pm$ 0.11 \\
\hline
\end{tabular}
\end{center}
\end{table}

A comparison of the gluon and quark factorial and cumulant moments
has also been performed by OPAL. Cumulants $K_q$ of order $q$
measure the genuine $q$-particle correlation. The predictions of
these quantities \cite{dremin} were only found to agree
``qualitatively'' with the data \cite{opal_vanc}. Partly this
seems understandable as the leading order (LO) , next to LO (NLO)
and NNLO predictions differ strongly. A recent
numerical calculation \cite{lupia} continuing the work of
\cite{ochs_lupia}, which now also considers energy conservation,
the correct low energy behaviour as well as OPAL's inclusive gluon jet
definition, describes the data well.

\begin{figure*}
\center
\parbox{8.6cm}{ \epsfig{figure=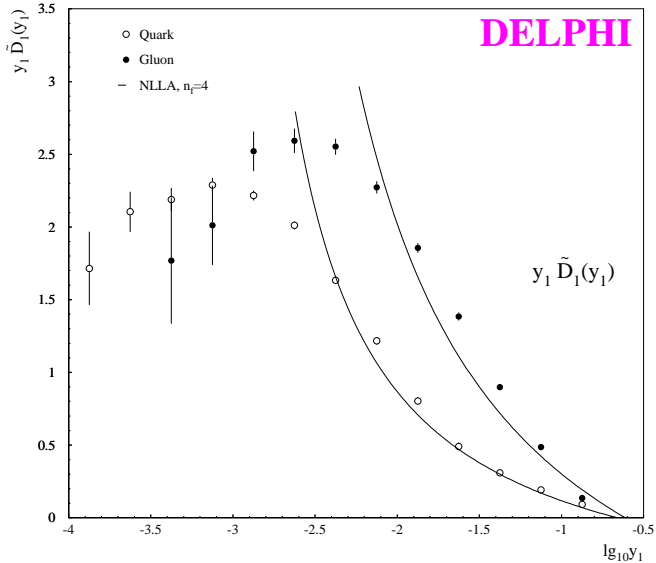,width=8.6cm} }
\hfill
\parbox{8.6cm}{ \epsfig{figure=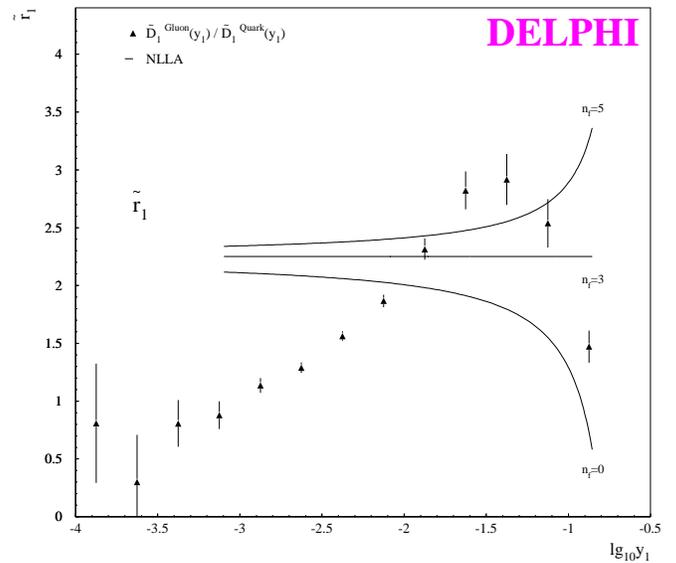,width=8.6cm} }
\caption{Gluon and quark splitting kernels (left) and their ratio
(right) as measured by DELPHI. Theoretical expectations are shown
as lines.
} \label{fig:d1tilde}
\end{figure*}

Also charged hadron spectra have been studied for the
inclusive gluons \cite{opal_vanc}, see Fig. \ref{fig:fragfct}.
Consistent with previous studies (see e.g. \cite{qg1_delphi}) a
roughly exponential falloff is observed for the distribution of
the scaled charged hadron energy, $x_E$, which is more pronounced
for gluons than for quarks. At large $x_E$ the gluon
distribution is suppressed with respect to the quark result by more than
one order of magnitude, i.e. the production of fast, so-called leading
particles is
strongly suppressed in gluon jets. At the smallest $x_E$, however,
the gluon to quark ratio reaches values of up to 1.8. The
distribution of the rapidity with respect to the sphericity axis of the
event shows a corresponding behaviour. This choice of the
observable ``zooms'' the range of low hadron energies. At the
smallest rapidities ($y<2$)  a
high gluon to quark ratio is observed as expected in \cite{catani}:
\begin{eqnarray}
\frac{dN^{gluon}/dy}{dN_{ch}^{quark}/dy} = 1.815 \pm 0.038 \pm 0.062
\label{eqn:opalratrat}
\end{eqnarray}
Although not mentioned in \cite{opal_vanc} it is important to note
that in the rapidity range $3 \leq y \leq 6$ the gluon
distribution is below the quark distribution.
The area between the
gluon and the quark distribution amounts to almost one unit of
charged multiplicity.
This suppression further extends down to rapidities of $y\sim 2$ as is
observed from Fig. \ref{fig:fragfct}d) and necessarily needs to be taken
into account when comparing the QCD expectation for the gluon to quark
multiplicity ratio to data, e.g. for Eqn. \ref{eqn:opalmultrat}.

In order to verify the QCD prediction of a stronger splitting
probability for gluons and also to clarify the origin of the small
observed multiplicity ratio DELPHI directly measured the so-called
splitting kernels. These are the most important piece of the
evolution equations for the generating functions or functionals
which describe the jet- and hadron multiplicities
(see e.g. \cite{khoze_ochs}). The importance of these kernels, commonly known
is the special (collinear) case of the Altarelli-Parisii kernels,
can be understood from the following simplified argumentation \cite{qg_splitt}.
Consider a given number of partons $N_1(y)$ which have not split (or decayed)
at a given jet resolution $y$. The number of partons which will
split in an interval $\Delta y$ then is expected to be:
\begin{eqnarray}
\Delta N_1(y) &=& -F(y) \cdot N_1(y) \cdot \Delta y ~~~\Longrightarrow
\nonumber \\
F(y) &=& -\frac{1}{N_1(y)} \cdot \frac{dN_1(y)}{dy}
\label{eqn:kernel}
\end{eqnarray}
This equation is analogous to the radiative decay equation,
however, with a $y$ dependent decay ``parameter'' $F(y)$ describing
the dynamics of the process. This is
the splitting kernel. Eqn. \ref{eqn:kernel} can be expressed by
the usual jet rates $R_n=N_n/N_{tot}$ and then yields:
\begin{eqnarray}
-F(y) = \tilde{D}_1(y) \approx \frac{d \log R_1(y)}{dy}
\label{eqn:kernel2}
\end{eqnarray}
In analogy to the differential jet rates known form the $\as$
analyses, the quantity $\tilde{D}_1(y)$ is called modified
differential 1-jet rate. It is a direct experimental measure of
the splitting kernel.

\begin{figure}
\center \epsfig{figure=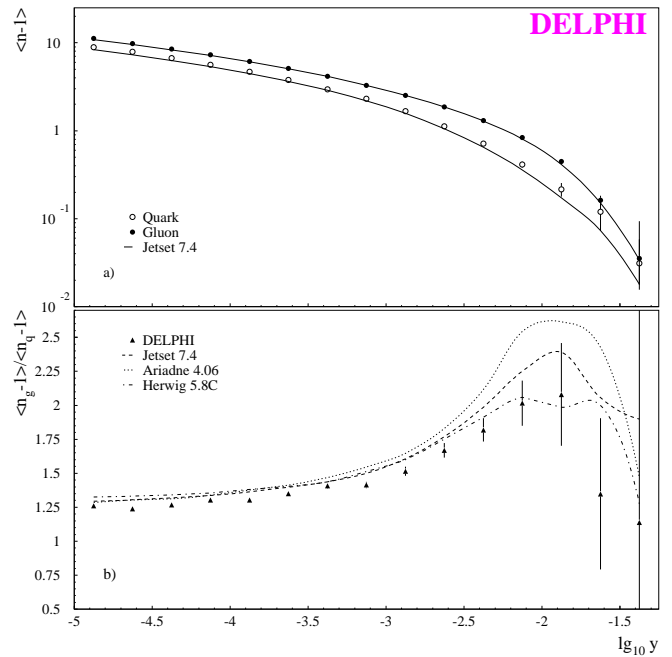,width=8.6cm} \caption{ Average
subjet multiplicity minus 1 for gluon and quark jets (upper plot)
and gluon to quark ratio (lower plot) as function of the resolution $y$. }
\label{fig:subjetmult}
\end{figure}
\begin{figure*}
\parbox{8.6cm}{ \epsfig{figure=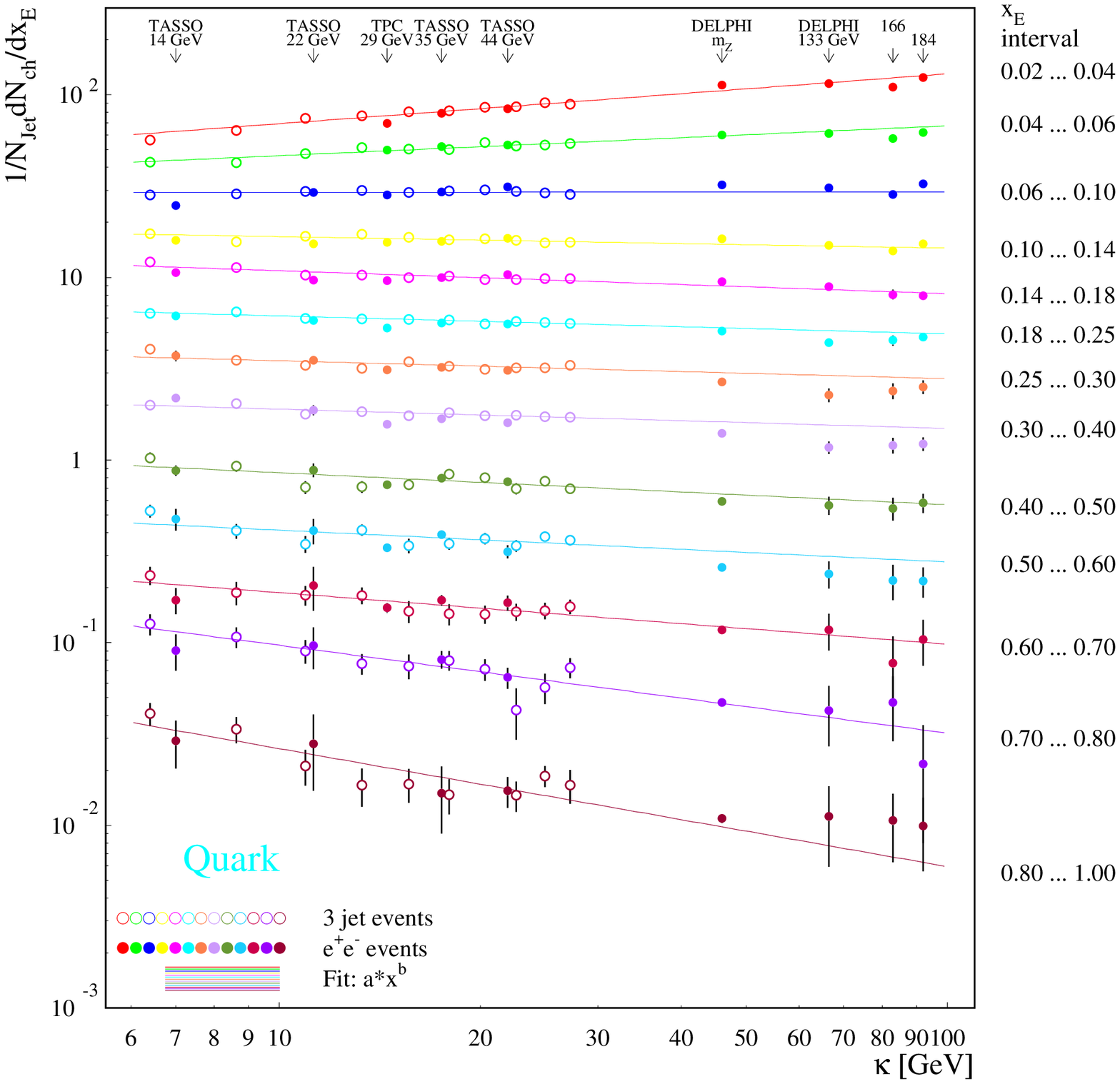,width=8.6cm} }
\hfill
\parbox{8.6cm}{ \epsfig{figure=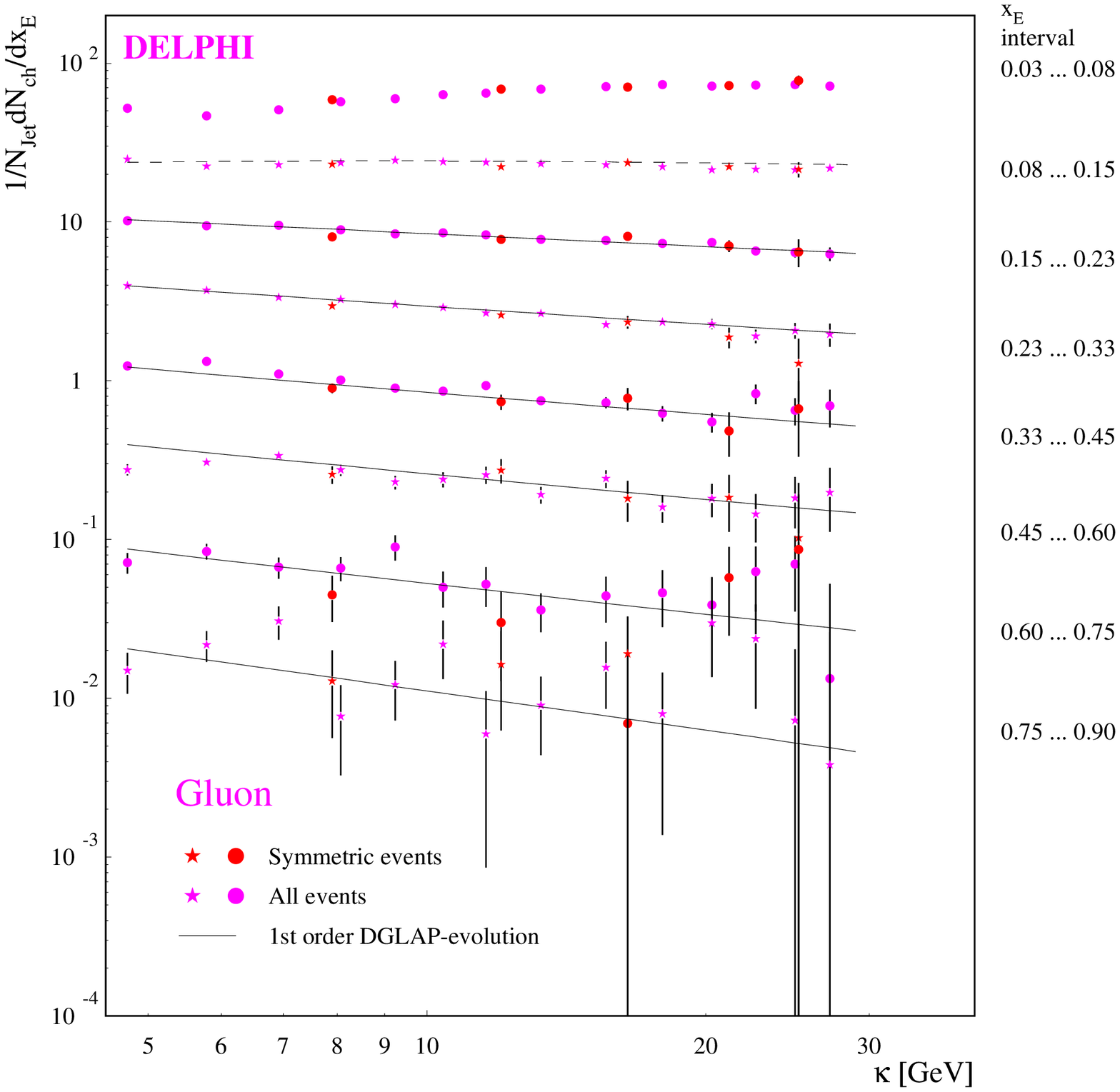,width=8.6cm} }

\center  \caption{ Left side: Scale dependence of the quark
fragmentation function measured in three-jet events as function of the
hardness scale $\kappa$ compared to measurements in \epem
annihilation. Lines are power fits to the three-jet data. Right side:
Scale dependence of the gluon fragmentation function measured in 3
jet events as function of the hardness scale $\kappa$. Data points
belong to symmetric and non-symmetric topologies. Lines result
from a simultaneous DGLAP fit of the gluon and quark fragmentation
function. } \label{fig:fanplot}
\end{figure*}

The result of the DELPHI measurement for gluons and quarks as
function of $y$ as defined with the Durham algorithm \cite{durham}
as well as the corresponding ratio is shown in Fig.
\ref{fig:d1tilde}. At large $y$ the data about follow the
perturbative expectation. A deviation at very large
$y$ (best seen in the gluon to quark ratio) is due to the structure of the
underlying three-jet events \cite{qg_splitt} (see also behaviour
of the subjet multiplicity-1 in \cite{aleph_splitting}).
The gluon
and the quark cannot be treated as independent objects in this limit
as the large $y$  splitting resolves the underlying event. At
smaller $y$ both the quark and the gluon data fall below the
perturbative expectation. This is to be expected due to
non-perturbative hadronization effects. It is most important, that
this influence sets in already at larger $y$ (i.e. ``earlier'') for gluons compared
to quarks. This causes the initially large gluon to quark ratio to
fall off rapidly with falling $y$.

In the framework of the HERWIG
cluster fragmentation model \cite{herwig} this effect can 
easily be understood. Quarks are valence particles of hadrons whereas
gluons are not. Gluons have to be split into a \qqbar\ -pair first.
So hadronisation effects should be visible already at larger $y$
in case of gluons. The same effect is responsible for the
suppression of the fragmentation function at large $x_E$ or
large rapidity $y$ as discussed above.

The behaviour of the so called sub-jet multiplicity
$<n_{jet}>=\sum_i i \cdot R_i$ (see Fig. \ref{fig:subjetmult}a))
illustrates that this is indeed the reason for the gluon to quark
multiplicity ratio to be smaller than expected. Initially at 
$\log{y}\approx-2$ the subjet multiplicity-1 ratio \footnote{1 is
subtracted to consider the initially present parton.} (see
Fig.\ref{fig:subjetmult}b) reaches about the expected value
$\approx$2. Then in coincidence with the deviation of
$\tilde{D}_1^g(y)$ from the perturbative expectation this ratio
falls off and rapidly reaches values typical for the hadronic
multiplicity ratio. Note from Fig. \ref{fig:subjetmult}a) that this
happens when the typical subjet multiplicities are around 3, thus
at much larger $y$ than most hadrons are formed.

\subsection{Scale Dependence of Quark
and Gluon Fragmentation}\label{subsec:gq_dyn}

DELPHI in several publications
\cite{qg_splitt,vanc_scaling,vanc_mult,qg1_delphi} attempted to
compare the scale dependence thus the dynamical evolution  of
gluons and quarks. Using the full data set containing about 85000
identified gluon jets a detailed comparison of the gluon and quark
fragmentation function $D(x_E)$ as function of $\kappa$ is
performed \cite{vanc_scaling}. The measurement is shown in Fig.
\ref{fig:fanplot}. The quark fragmentation function as obtained
from quark jets in three-jet events (left side) was fitted by a power
ansatz superimposed to the data (open points). Also shown (full
points) is the fragmentation function as measured in \epem
annihilation from PETRA energies \cite{frag_tasso} up to the highest
LEP energies. Both measurements agree very well, enforcing the
interpretation of $\kappa$ as a valid scale for jet evolution in
three-jet events. It should, however, be mentioned that the
transverse momentum of a jet with respect to the event axis works similarly
well. The corresponding result for gluons is shown on the right
side of Fig. \ref{fig:fanplot}. Results for symmetric and for all
topologies agree well, strengthening again the validity of the
analysis and the scale. The fits shown with the scale dependence
of the gluon fragmentation function result from a simultaneous
$\rm 1^{st}$ order DGLAP evolution together with the quark
fragmentation function. The strong coupling, the colour factor
ratio $C_A/C_F$, as well as the parameters of the analytic ansatz
of the fragmentation functions were treated as free parameters. The
fit yields the following preliminary result for the colour factor
ratio:
\begin{eqnarray}
\frac{C_A}{C_F}=2.44 \pm 0.21 (stat.)
\end{eqnarray}

\begin{figure}
\center
\epsfig{figure=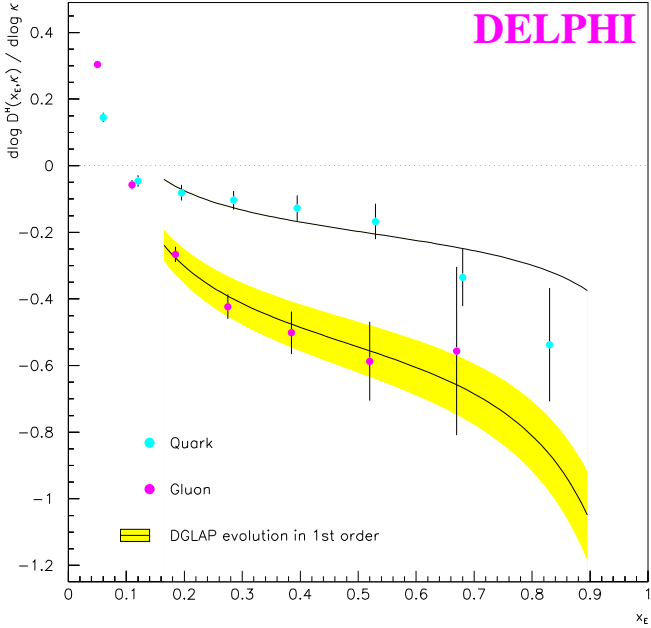,width=8.6cm} \caption{ Slopes
of the scale dependence of the quark and gluon fragmentation
function as function of $x_E$. The shaded area indicates the stat.
uncertainty of $C_A$ from the DGLAP fit. Data points result from
power law fits of the individual  $x_E$ intervals. }
\label{fig:qgD_slope}
\end{figure}

Fig. \ref{fig:qgD_slope} compares the logarithmic slope of the gluon and quark
fragmentation function as function of $x_E$.
The function is the result of the DGLAP fit of the data. The shaded
area indicates the uncertainty on $C_A$ for fixed $C_F=4/3$.
The data points were obtained from power law fits to each gluon and quark $x_E$
interval individually.
As expected from the structure of the DGLAP equation scaling
violations are much stronger for gluons than for quarks
\footnote{Note that the Altarelli-Parisii
splitting functions include the individual
colour factors multiplicatively.}.
Due to the stronger scaling violations in gluon jets
\ppbar\ sscattering which is dominated by gluon-gluon scattering should
be well suited to measure $\as$ form the scaling violation of the
charged hadron fragmentation function.

\begin{figure*}[t]
\center
\parbox{8.6cm}{ \epsfig{figure=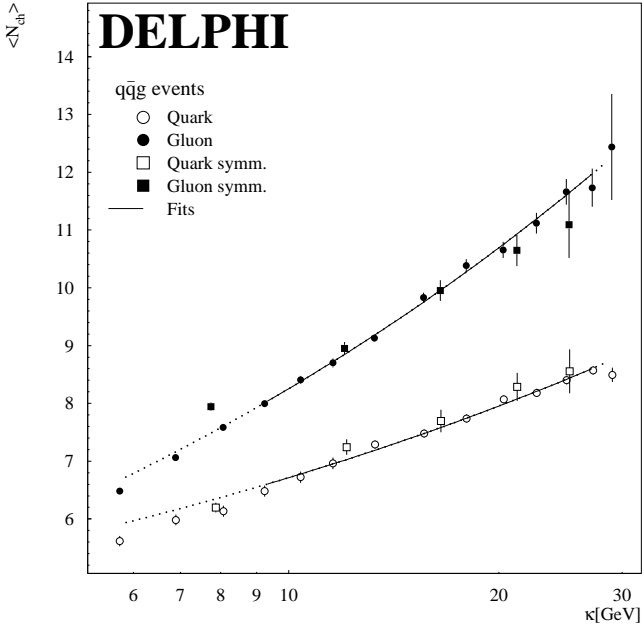,width=8.6cm} }
\hfill
\parbox{8.6cm}{ \epsfig{figure=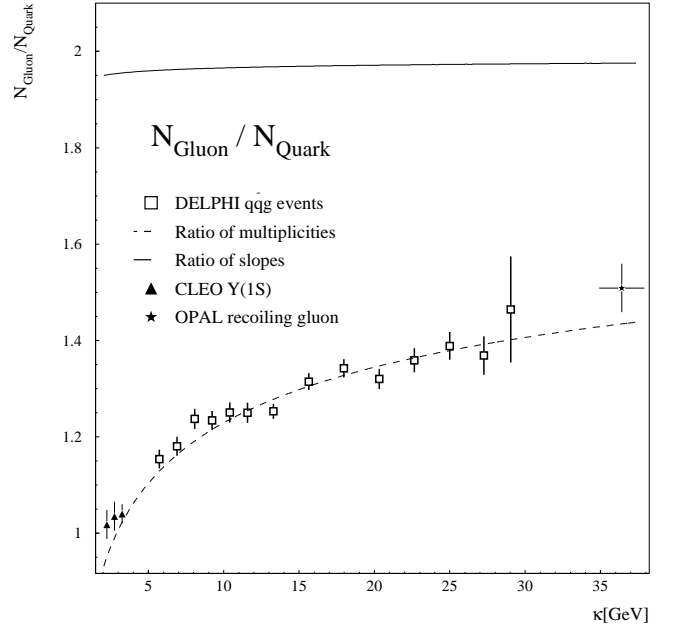,width=8.6cm} }
\caption{
Left: Scale ($\kappa$) dependence of the jet multiplicity in gluon and
quark jets as measured by DELPHI.
Lines are the result of a fit of Eqn. \ref{mult_slope}
Right: Ratio of the gluon to quark multiplicity as function of the
scale.
The lower line is the ratio of the fits on the right.
The upper line is the ratio of the slopes of these fits.
}
\label{fig:mult_scale}
\end{figure*}

It is evident from the gluon to quark comparison at low $x_E$ - indicating a
ratio of about 2 (see Fig. \ref{fig:qgD_slope}) - that the scale dependence of the hadron
multiplicity opens a new way in testing the prediction:
\begin{eqnarray}
\frac
{\left<N_{ch}^{gluon} \right>}
{\left<N_{ch}^{quark} \right> } \cong \frac{C_A}{C_F}
\label{multrat_exp}
\end{eqnarray}
insensitive to the non-perturbative effects discussed above.
Note that the prediction Eqn. \ref{multrat_exp} is expected to be valid for any
large scale if non-perturbative effects can be neglected.
Consequently also the ratio of the derivative of the
multiplicities with respect to the relevant scale (here taken to
be $\kappa$) is equal to the colour factor ratio:
\begin{eqnarray}
\frac{\partial < N_{ch}^{gluon}(\kappa) >/\partial \kappa}
{\partial < N_{ch}^{quark}(\kappa) >/\partial \kappa} =
\frac{C_A}{C_F}
\label{mult_slope}
\end{eqnarray}
This prediction is equivalent to Eqn. \ref{multrat_exp} but, as the
multiplicity evolves slowly with scale, should hold already at
much smaller scale. The comparison of the gluon and quark splitting kernels as
well as the behaviour of the fragmentation function suggest that
most of the non-perturbative difference between gluons and quarks
is due to the fragmentation of the leading quark or gluon,  thus
is to be expected to happen at very small scale. A corresponding
calculation in the framework of the colour dipole model
\cite{eden} confirms this expectation.

Fig. \ref{fig:mult_scale}
compares the dependence of the multiplicity assigned to gluon and
quark jets as function of $\log \kappa$. It is immediately evident
from this plot that the slope of the gluons is roughly twice as
big as for quarks thus already impressingly confirming the QCD
expectation Eqn. \ref{mult_slope}. The data was fit with the
predictions for the evolution of the multiplicity with scale as
given in \cite{DKT,webber}, however, adding constant terms to
account for non-perturbative differences between gluons and
quarks, thus:
\begin{eqnarray}
\nonumber
<N>_{ch}^{q}(\kappa)&=&<N>_{ch}^{perturb.}(\kappa) \cdot + N_{0}^{q}\\
<N>_{ch}^{g}(\kappa)&=&<N>_{ch}^{perturb.}(\kappa) \cdot
r_{gq}(\kappa) + N_{0}^{g}
\label{eqn:mult_fit}
\end{eqnarray}
$r_{gq}(\kappa)$ is a NNLO prediction for the multiplicity ratio
as given in \cite{gaffney_mueller}.
The ansatz represents the data well. The resulting colour factor ratio
is:
\begin{eqnarray}
\frac{C_A}{C_F}=2.14 \pm 0.10 (stat.)
\end{eqnarray}
No systematic error is given with this measurement as a
superior method is discussed also
in \cite{vanc_mult} which determines the colour factor ratio from
the scale dependence of the multiplicity of three-jet events. It
is insensitive to ambiguities induced by the choice of the scale
and also accounts for coherent radiation
from a \qqbarg\ -ensemble (see also \cite{stroehmer}). The
non-perturbative term introduced in this study which accounts for
differences in the fragmentation of the primary quark or gluon
is $N_{0}^{q} - N_{0}^{g} = 1.97 \pm 0.03 \pm 0.33$.

It is instructive to estimate a lower limit for
$N_{0}^{q} - N_{0}^{g}$ from the gluon and quark
fragmentation functions \cite{qg_splitt} in the momentum range where
the gluon is below the quark fragmentation function (compare also
Fig.\ref{fig:fragfct}a,c)).
This limit can be calculated by subtracting the observed gluon multiplicity from the
(wrongly) expected one: $N_g^{expected}\approx 2 N_q$.
This yields $N_{0}^{q} - N_{0}^{g} > 1.22 \pm 0.08 (> 1.15 \pm 0.2 )$ from $Y$
and {\em Mercedes} events, respectively.
As the suppression persists to lower $x_E$ or $y$ (see Fig. \ref{fig:fragfct}b,d))
it probably amounts for
the full non-perturbative term $N_{0}^{q} - N_{0}^{g}$.

The left side of Fig. \ref{fig:mult_scale} presents the gluon to quark
multiplicity ratio, the ratio of the fitted functions
Eqn. \ref{eqn:mult_fit} as well as the corresponding slope ratio. Also
included are measurements at very small scale (4-7~GeV) performed
by CLEO \cite{cleo} in $\Upsilon (1S) \rightarrow {\rm gg}\gamma$
decays and the OPAL result
for inclusive gluons. For this point $\kappa$ has been
estimated from the average gluon energy and the angle cuts given
in \cite{opal_vanc}.
As this result compares gluon to pure light quark (uds) jets it
may be expected to lie slightly higher than the extrapolation of the DELPHI result.
The fit describes all measurements well,
confirming that the non-perturbative effects can be absorbed   effectively
into constant terms allocated to the leading particles in a quark
or gluon jet.



\subsection{Reconstruction of Colour Connections \label{subs:connect}}
The stronger radiation off a gluon may also be interpreted as due
to two colour connections compared to one in the case of quarks.
One may then try to reconstruct the colour connections in an event
to distinguish e.g. 4 quark final states (like $\rm WW \rightarrow
q\bar{q}q\bar{q}$ events, 2 connections) from QCD background ($\rm
q\bar{q}gg$, 3 connections). One may also try to reduce the
combinatorial background in combining only the colour connected
\qqbar\ -pairs in WW events. Besides several theoretical proposals related to
this subject \cite{theo_flow} now first experimental results are
available \cite{delphi_3jet,delphi_ww}. These are based on the idea of
reconstructing the colour portrait of a process \cite{orava}. To
introduce the algorithm recall first the structure of the soft
hadron (labeled $h$) production cross-section in a hard $n$ parton
(labeled $i,j$) initial state \cite{string,qcd_and_collider}:
\begin{eqnarray}
\nonumber
d\sigma_h^{soft} &\propto& d\sigma^{hard} \cdot \alpha_s \cdot
d\Omega \cdot \frac{dE_h}{E_h} \sum_{i,j~hard}^n C_{ij} \cdot
W_{ij}\\ W_{ij}&=&\frac{sin^2\theta_{ij}/2}{sin^2\theta_{ih}/2
\cdot sin^2\theta_{jh}/2}
\end{eqnarray}
Important pieces in this equation are the colour factors $C_{ij}$
which contain the information about the colour connections
($C_{ij}=0$ if no connection) and the so called antenna pattern
$W_{ij}$ which determine the angular structure of the radiation.
The poles $\theta_{kh} \rightarrow 0$ appear in the direction of
the hard partons and lead to the formation of the jets. The
nominator contains the angular dependence known from the hardness
scale. The idea is to reconstruct the colour connections by
adequately weighting the observed soft hadron production. The
algorithm proceeds as follows:
\begin{itemize}
\item
Reconstruct jet directions from leading particles.
\item
Calculate the probability for un-associated (low energy, large angle)
particles to stem from a cluster (hard parton) j:
\begin{equation}
w_{hj}=\frac{C_j}{k^2_{hj}}=\frac{C_j}{2E^2_h \sin^2\theta_{hj}/2} ~~~.
\label{eqn:stem_prob}
\end{equation}
Renormalise so that $\sum w_{hj} = 1$.
\item
Associate hadron $h$ to the cluster pair which maximises
$w_{hkl}=w_{hl}+w_{hk}$.
Calculate the colour connection coefficient:
\begin{equation}
W_{kl}=C \sum_{h~assoc.} g(E_h) \cdot w_{hkl}~~~.
\nonumber
\end{equation}
$g(E_i)$ is an energy dependent weight, $C$ is chosen so to
properly normalise the $W_{kl}$; $\sum W_{kl}=2$.
\end{itemize}

\begin{figure}[t]
\center \epsfig{figure=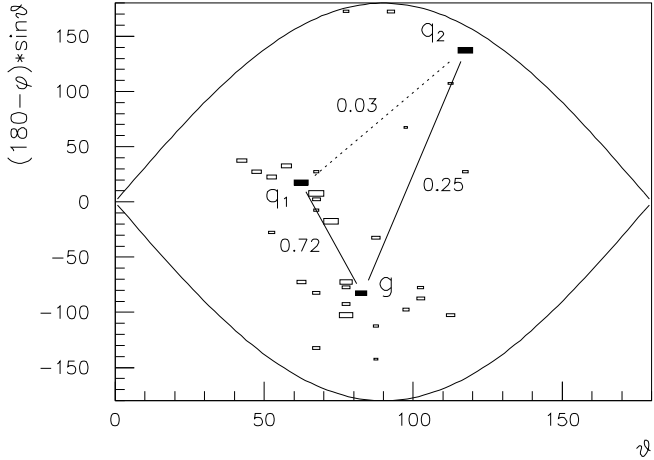,width=8.6cm} \caption{ Fish-eye
plot of the hadron distribution in a typical {\em Mercedes} event
(Monte Carlo).
} \label{fig:fisheye}
\end{figure}

\begin{figure}[t]
\center \epsfig{figure=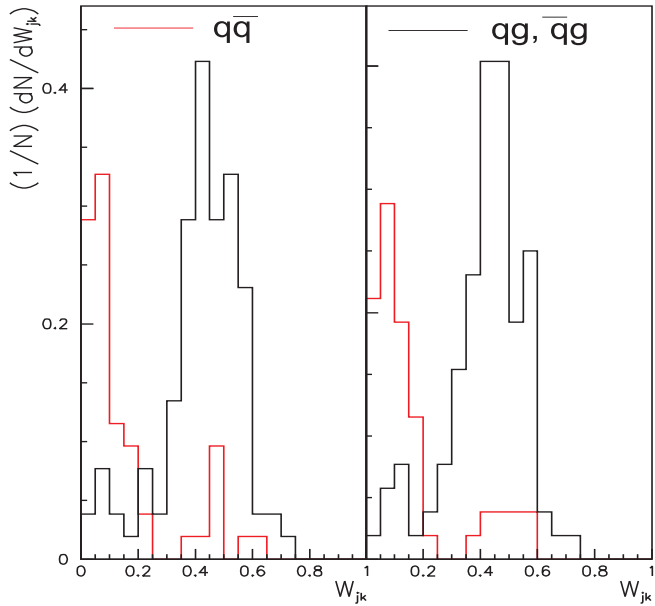,width=8.6cm,height=8.cm}
\caption{ Weight
distribution for {\em Mercedes} events (DELPHI data).
} \label{fig:merc_weights}
\end{figure}

To test the technique it has been applied to {\em Mercedes} events
(Monte Carlo and data) where the (heavy) quark jets were
additionally identified using an impact parameter tag \cite{delphi_3jet}.
Fig.\ref{fig:fisheye} shows a typical event. It is clearly observed
that the soft hadrons (boxes) accompany the quark-gluon
colour connections indicated by the lines. The corresponding colour
connection coefficients clearly distinguish the $\rm qg (\bar{q}g)$ from the
\qqbar\ case. Fig. \ref{fig:merc_weights} (left) shows the colour
connection coefficients which were obtained from DELPHI events
of the {\em Mercedes} topology. The additional impact parameter
identification allows to measure directly the purity and
efficiency of the algorithm. The data show that purities of 90\%
at 80\% efficiency can be obtained.

\begin{figure}[h]
\center \epsfig{figure=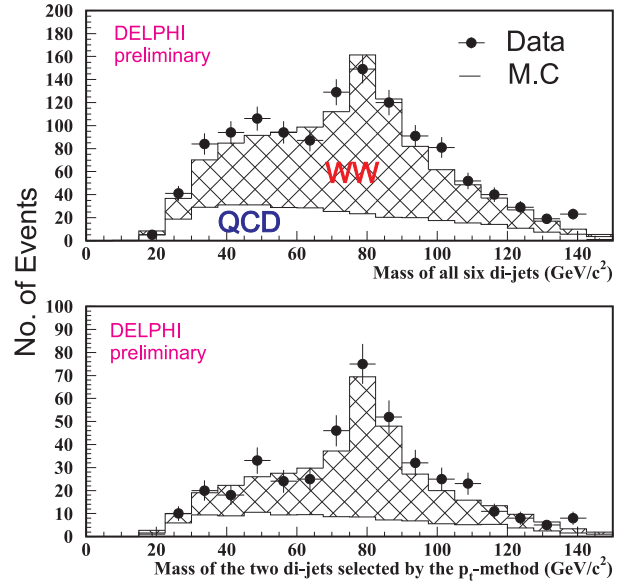,width=8.cm}
\caption{ Jet-jet
mass distribution for four-jet events measured by DELPHI at 189 GeV.
The upper plot contains all combinations, the lower only those
selected with the algorithm described in the text. Histograms
indicate the Monte Carlo predictions for WW and QCD events. }
\label{fig:ww_189}
\end{figure}

A modification of the algorithm has also been applied to the
analysis of $\rm WW \rightarrow q\bar{q}q'\bar{q'}$ events at 189 GeV
centre-of-mass energy measured by DELPHI. As proposed in
\cite{norrbin_sjoestrand} all (hypothetical) strings connecting
the 4 parton final state were boosted to their rest systems. Then
the transverse momentum of the hadrons with respect to the string was
calculated and similar to Eqn. \ref{eqn:stem_prob} weights were
defined using the transverse momentum of the hadrons
$w_{hj}=C_h/p_t^h$. The string configuration which minimises the
sum of the transverse momenta was selected. Fig. \ref{fig:ww_189}
shows the mass spectrum of all possible jet-jet pairs (upper plot)
and taking only the combinations selected by the algorithm
outlined above. A clear reduction of the combinatorial background
due to wrong $\rm q\bar{q}'$ pairing is indicated. Contrary to
kinematic cuts the method does not bias the W-mass. So far no
attempt has been made to reject QCD background. Thus the appearend
reduction of this contribution is due to the reduced combinatoric
only. Currently the method can only be used as an additional tool
as the efficiency is only about 50\%. Improvements are however to
be expected in the near future.

\section{Summary}\label{sec:summary}
The introduction of energy scales in the comparison of gluon and
quark jets opened the possibility for studies of the scale
dependence of gluon and quark jet properties.
This in turn lead to the
understanding of the (smaller than originally expected) gluon to quark
multiplicity ratio due to non-perturbative effects in the
fragmentation of the leading gluon or quark
and to the possibility to really measure the
colour factor ratio from a direct gluon to quark jet comparison.
These observations will soon be further exploited in order to identify
quark and gluon jets or colour connections in multi-jet
environments.

\section*{Acknowledgements}
I would like to express my gratitude to the colleagues who
provided me with input for this talk: B. Gary, A. Kiiskinen, S.
Lupia, V. Nomokonov and R. Orava.
Especially  I like to thank O. Klapp, P.
Langefeld and M. Siebel for their persistent effort and the
pleasant time during the gluon analysis of the Wuppertal group of DELPHI.

\end{document}